\documentclass[12pt]{article}

\usepackage{graphicx}
\usepackage{psfrag}
\usepackage{flafter}
\usepackage{amssymb}

\newcommand{\non}{\nonumber}
\def\be{\begin{equation}}
\def\ee{\end{equation}}
\def\bea{\begin{eqnarray}}
\def\eea{\end{eqnarray}}

\def\pa{\partial}                             
\def\sbra#1{\left\langle #1\right|}           
\def\sket#1{\left| #1\right\rangle}           
\def\sVEV#1{\left\langle #1\right\rangle}     
\def\ha{\frac12}                              
\def\pder#1#2{{\pa #1\over\pa #2}}            

\def\a{\alpha}

\def\d{\delta}

\def\h{\eta}

\def\p{\pi}           
\def\s{\sigma}        
\def\t{\tau}

\def\co{{\cal O}}
\def\cp{{\cal P}}

\begin{document}

\title{{\bf Stochastic Simulation of Grover's Algorithm}}

\author{L.R.U. Manssur, R. Portugal \\
{\small Coordena\c{c}\~{a}o de Ci\^encia da Computa\c{c}\~{a}o}\\
{\small Laborat\'{o}rio Nacional de Computa\c{c}\~{a}o
Cient\'{\i}fica - LNCC}\\ {\small Av. Get\'{u}lio Vargas 333,
Petr\'{o}polis, RJ, 25651-070, Brasil}\\ \textit{e-mail:
\{leon,portugal\}@lncc.br} } \maketitle

\begin{abstract}

We simulate Grover's algorithm in a classical computer by means of
a stochastic method using the Hubbard-Stratonovich decomposition of
$n$-qubit gates into one-qubit gates integrated over auxiliary
fields. The problem reduces to finding the fixed points of the
associated system of Langevin differential equations. The equations
are obtained automatically for any number of qubits by employing a
computer algebra program. We present the numerical results of the
simulation for a small search space.

\end{abstract}

\section{Introduction}

Grover's and Shor's algorithms \cite{grover1,shor} are landmarks
in the development of Quantum Computation since the foundation of
this area in the 1980's \cite{benioff,feynman,deutsch}. Shor's
algorithm provides an exponential speed up over currently known
classical factorization algorithms. Although a classical
polynomial factorization algorithm might be found, it is very
unlikely that its complexity would be as low as Shor's $O(n^3)$.
Compare with the polynomial $O(n^{12}\mbox{ poly}(\log n))$
algorithm for the much easier problem of primality test recently
developed by Agrawal, Kayal, and Saxena \cite{aks}. Therefore, it
is unlikely that a polynomial factorization algorithm, if it
exists, would have complexity under $O(n^{12})$.

Shor's algorithm is the best candidate to be simulated in
classical computers, since the price paid by the simulation may
still be low compared to the exponential timing of classical
algorithms. Unfortunately, the techniques used in this paper do
not seem to be suitable for Shor's algorithm \cite{cerf}. 

Grover's algorithm plays an important role in Quantum Computation,
since it provides a proof that quantum computers are faster than
classical ones for unstructured database searching. It has
complexity $O(\sqrt N)$ while the best classical algorithm has
complexity $O(N)$, where $N$ is the database size. Grover's
algorithm is optimal using oracle searching  \cite{bennett,boyer},
therefore the quadratic speed up is the best one can achieve in
this case. The hope in simulating it in classical computers with a
better efficiency compared to the classical algorithm is very low.
Despite this fact, Grover's algorithm provides a laboratory to
test simulation techniques that might shed new light in
distinguishing the way classical and quantum computers work.

In a seminal paper, Feynman \cite{feynman} argues that classical
computers can simulate quantum ones only with an exponential slow
down. It is easy to see that an exact simulation uses either an
exponential amount of memory or takes an exponential time to
accomplish generic tasks that could be done by a quantum computer.
A generic quantum state of $n$ qubits has $2^n$ complex amplitudes
which must be taken into account by the classical computer. An
alternative option is a stochastic simulation by some
``non-deterministic computer''. The central issue is the local
nature of classical computers. A classical computer of the same
physical size of a quantum computer working under local classical
physical laws does obey the Bell inequality \cite{CHSH} and cannot
reproduce the same results of a computer which follows the quantum
mechanical laws.

Feynman's arguments, on the other hand, are not quantitative. The
question is: what is exactly the loss introduced by the stochastic
simulation, in such a way that the result given by the quantum
computer is still reproduced?

An interesting stochastic simulation technique in the quantum
computation context was introduced by Cerf and Koonin \cite{cerf}.
The quantum computer is viewed as a many-body dynamics which can
be reduced to the time evolution of single qubits integrated in
auxiliary fields. They used the Hubbard-Stratonovich
\cite{alhassid,kashiwa} representation to obtain an expression for
two-qubit gates in terms of two one-qubit gates with two auxiliary
fields . They used a Monte Carlo Method \cite{johnson,heinrich}(in
fact the Langevin equations with equivalent solutions
\cite{okano,parisi}) to reduce the exponential size of the
associated Hilbert space to an equivalent coupled system of
differential equations with a polynomial number of auxiliary
fields.

It is known that a general quantum circuit can be decomposed in
terms of one and two-qubit gates, called universal gates
\cite{barenco,chuang}. On the other hand, it is also known that
this decomposition may use an exponentially large number of
universal gates \cite{chuang} (and auxiliary fields). In that
case, the gain provided by the stochastic method would be lost in
the decomposition in terms of the universal gates.

In this paper, we generalize Cerf and Koonin's method to general
$n$-qubit gates, avoiding the decomposition of $n$-qubit gates
into universal gates. The generalization is straightforward and
welcome in simulating Grover's algorithm for a general number of
qubits. We apply the method for Grover's algorithm, and make use
of a Maple program that generates automatically the associated
system of Langevin equations. The equations are solved numerically
(in Maple and in C) giving the relaxation values of the auxiliary
fields which correspond to the fixed points of the Langevin
dynamics. We present the numerical results for the simplest case
of Grover's algorithm and discuss
some problems we are facing with the method.

\section{Decomposition in terms of one-qubit gates}

Consider a quantum circuit of $g$ gates ($U_k, k=1,...,g$) in a
$n$-qubit quantum computer. Each gate $U_k$ acts on two or more
qubits $(j_k^{(1)},j_k^{(2)},...,j_k^{(n_k)})$. The qubit indices
within a gate $U_k$ are put in parentheses. They do not coincide
necessarily with the overall qubit indices, which will be denoted
in square brackets.

In order to clarify the notation consider the following example:
if $U_k$ is a two-qubit gate acting on the third and fifth qubits,
then $n_k=2$, $j_k^{(1)} = 3$ and $j_k^{(2)} = 5$. The index $s$
of $j_k^{(s)}$ runs from 1 to $n_k$. The actual values of
$j_k^{(s)}$ are in increasing order but not necessarily
consecutive with respect to the overall qubits, as shown in the
example. One-qubit gates will be trivially introduced in the next
section, so they will not be considered for a while.

The computation as a whole is performed by the unitary
operator $U$ given by
\be
\label{U}
U = \prod_{k=g}^1 U_k = U_g \; ... \; U_1
\end{equation}
where the product is in reverse order, so the operators act
in ascending order
from left to right on input kets. $U_k$ is a general $n_k$-qubit
gate, with $2 \leq n_k \leq n$. We assume that $U_k$ can be
decomposed as an exponential of a tensor product of one-qubit
gates, in the form
\be
\label{Uk}
U_k = e^{-i \a_k A_k^{(1)} \otimes A_k^{(2)} \otimes \; \dots \;
\otimes A_k^{(n_k)} },
\end{equation}
where $A_k^{(s)}$ acts only on qubit $j_k^{(s)}$. This equation
generalizes eq.(2) of Cerf and Koonin's paper \cite{cerf}.

The Hubbard-Stratonovich decomposition \cite{alhassid,kashiwa} for
a $n_k$-qubit gate follows from
\bea
\label{dec}
e^{-i \a_k A_k^{(1)} \otimes \; \dots \;
\otimes A_k^{(s)} } & = &  \int_{- \infty}^\infty d\s_k^{(s)} \;
e^{-i \a_k A_k^{(1)} \otimes \; \dots \; \otimes A_k^{(s-1)}
\otimes \s_k^{(s)} I_k^{(s)} } \times \non \\
& & \;\;\;\;\;\;\;\;\;\; \d(I_k^{(1)}\otimes \; \dots \;
\otimes I_k^{(s-1)} \otimes (A_k^{(s)} - \s_k^{(s)} I_k^{(s)}) ),
\eea
where $I_k^{(s)}$ is the identity matrix acting on the qubit
$j_k^{(s)}$, and from the following representation for the Dirac
delta function
\bea
\label{dec2}
\lefteqn{\d(I_k^{(1)}\otimes \; \dots \; \otimes I_k^{(s-1)}
\otimes (A_k^{(s)} - \s_k^{(s)} I_k^{(s)}) ) = } \hspace{15ex}
\non \\
& & \frac{1}{2 \p}
\int_{- \infty}^\infty d \t_k^{(s)} \; \a_k \; e^{-i \a_k \t_k^{(s)}
I_k^{(1)}\otimes \; \dots \; \otimes I_k^{(s-1)}
\otimes (A_k^{(s)} - \s_k^{(s)} I_k^{(s)}) }.
\eea
Applying decomposition (\ref{dec}) and (\ref{dec2}) recursively
$n_k-1$ times in eq.(\ref{Uk}), we introduce real auxiliary fields
$\s_k^{(s)}$ and $\t_k^{(s)}$, $2 \leq s \leq n_k$. For
simplicity, we will drop the identity matrices from now on. The
result is the following expression:
\bea
 U_k &= &\int_{- \infty}^\infty d \s_k^{(2)} \; d \t_k^{(2)}
\dots \; d \s_k^{(n_k)} d \t_k^{(n_k)} \; \left( \frac{\a_k}{2\p} \right)^{n_k-1}
e^{-i \a_k \s_k^{(2)}
\dots \; \s_k^{(n_k)} \; A_k^{(1)} } \; \times   \non \\
& &  \hspace{1.4in} e^{-i \a_k \t_k^{(2)} (A_k^{(2)} -
\s_k^{(2)} ) } \; \dots  e^{-i \a_k \t_k^{(n_k)}
( A_k^{(n_k)} - \s_k^{(n_k)} )}.
\eea
Substituting $U_k$ into eq.(\ref{U}) yields
\bea
U & = & \int_{- \infty}^\infty \prod_{k=g}^1 \left\{ d \s_k^{(2)}
\; d \t_k^{(2)} \dots \; d \s_k^{(n_k)} d \t_k^{(n_k)} \left(
\frac{\a_k}{2\p} \right)^{n_k-1} e^{i\a_k \sum_{s=2}^{n_k}
\s_k^{(s)} \t_k^{(s)} } \right. \non \\
 & & \hspace{1.7in} V_k^{(1)}(\s_k) V_k^{(2)}(\t_k^{(2)})
\; \dots \; V_k^{(n_k)} ( \t_k^{(n_k)}) \biggm\} ,
\eea
where
\bea
V_k^{(1)} (\s_k) & = & V_k^{(1)}(\s_k^{(2)},\dots, \s_k^{(n_k)})
= e^{-i \a_k \s_k^{(2)} \; \dots \; \s_k^{(n_k)} A_k^{(1)} }
\non \\
V_k^{(s)} ( \t_k^{(s)} ) & = & e^{-i \a_k \t_k{(s)} A_k^{(s)} },
\;\;\; 2 \leq m \leq n_k.
\eea
$\s_k$ denotes all the $\s_k^{(s)}$ for a given $k$, and the
superscript $(s)$ indicates in which qubit $A_k^{(s)}$ acts.

Up to this point, we have been considering $U$ as a composition of
$g$ gates. Using the Hubbard-Stratonovich representation, each
gate was decomposed into one-qubit gates, paying the price of
introducing $2(n_k-1)$ scalar fields and respective integrals. Now
we focus on the world line of each qubit and multiply
(horizontally) all one-qubit gates acting on that qubit. We end up
with $n$ one-qubit gates which are tensored out and integrated on
the auxiliary fields, yielding the following expression for $U$
\be
\label{UU}
U = \int D\s \; D\t \; e^{i \sum_{k=1}^g \a_k \sum_{s=2}^{n_k}
\s_k^{(s)} \t_k^{(s)} } \bigotimes_{l=1}^n U^{[l]}(\s,\t)
\end{equation}
where square brackets denote an overall qubit label (as
opposed to internal gate labels $j_k^{(s)}$), with
\be
\label{Ul}
U^{[l]}(\s,\t) = \prod_{k=g}^1 U_k^{[l]}(\s,\t),
\end{equation}
(here $\s$ and $\t$ denote all of them) and
\be
\label{Ukl}
U_k^{[l]}(\s_k, \t_k) = \left\{
\begin{array}{cc}
V_k^{(1)} (\s_k) & \mbox{if } l=j_1 \\
V_k^{(2)} (\t_k^{(2)}) & \mbox{if } l=j_2 \\
\vdots & \vdots \\
V_k^{(n_k)} (\t_k^{(n_k)}) & \mbox{if } l=j_{n_k} \\
I_k^{[l]} & \mbox{otherwise.}
\end{array}
\right.
\end{equation}

In the remaining of this section, we follow closely Cerf
and Koonin's approach \cite{cerf}. The last step of the
computation is the measurement of the first $m$ qubits,
$m \leq n$. We assume that the observable is the direct
product of $m$ one-qubit observables,
\be
\co = \bigotimes_{l=1}^{m} \co^{[l]}.
\end{equation}
>From the remaining qubits, $p \leq n-m$ have prescribed
values $\p_1,\p_2,\dots,\p_p$, so we define the
projector $\cp$ as
\be
\label{P}
\cp = \prod_{k=1}^p \cp_k,
\end{equation}
where $\cp_k = \sket{\p_k} \sbra{\p_k}$. The expectation value
of the observable $\co$ is given by
\be
\label{O}
\sVEV{\co} = \frac{\sbra{0_1 \dots 0_n} U^\dagger \co P U
\sket{0_1 \dots 0_n}}{\sbra{0_1 \dots 0_n} U^\dagger P U
\sket{0_1 \dots 0_n}}
\end{equation}
Using eqs.(\ref{UU}), (\ref{P}) and (\ref{O}) we get
\be
\sVEV{\co} = \frac{ \int D\s \; D\t \; D\s^\prime \; D \t^\prime \;
\exp( -i S(\s,\t,\s^\prime,\t^\prime)) \; \co(\s,\t,\s^\prime,\t^\prime)}
{\int D\s \; D\t \; D\s^\prime \; D \t^\prime \;
\exp( -i S(\s,\t,\s^\prime,\t^\prime)) }
\end{equation}
where
\bea
\label{S}
S(\s,\t,\s^\prime,\t^\prime) & = & - \sum_{k=1}^g \a_k
\left( \sum_{s=2}^{n_k} \s_k^{(s)} \t_k^{(s)}  -
\s_k^{\prime (s)}\t_k^{\prime(s)} \right) \non \\
& & + i \sum_{l=1}^{n} \ln \sbra{0_l}
U^{\dagger[l]}(\s^\prime, \t^\prime) P^{[l]}
U^{[l]}(\s, \t) \sket{0_l}
\eea
and
\be
\label{O}
\co(\s,\t,\s^\prime,\t^\prime) = \prod_{l=1}^n \frac{\sbra{0_l}
U^{\dagger[l]}(\s^\prime, \t^\prime) \co^{[l]}
P^{[l]} U^{[l]}(\s, \t) \sket{0_l} } {\sbra{0_l}
U^{\dagger[l]}(\s^\prime, \t^\prime)  P^{[l]}
U^{[l]}(\s, \t) \sket{0_l}  }
\end{equation}
where $\co^{[l]} = I$ for $l>m$.

The most promising method to calculate $\sVEV \co$ is by using the
associated Langevin equations given by
\be
\label{langevin}
\frac{d \s^{(s)}_k}{dt\;} = - \frac i2 \pder {S\;} {\s^{(s)}_k}
+ \h^{(s)}_k(t)
\end{equation}
where $t$ is a simulation time, $\h^{(s)}_k$ is a real Gaussian
white noise and the scalar fields have been extended to the
complex plane. The stochastic estimate of $\sVEV \co$ is
\be
\sVEV{\co} \approx \frac 1T \int^{t+T}_t dt \co(\s(t), \s^\prime (t))
\end{equation}
where $\s$ and $\s^\prime$ are solutions of eq.(\ref{langevin}), and $T$
represents a sufficiently large simulation time. In the next section, we
will obtain that $\s$ and $\s^\prime$ converge to some final limiting value,
so we will use the formula
\be
\sVEV{\co} \approx \co(\s_{\mbox{\scriptsize final}},
\s^\prime_{\mbox{\scriptsize final}}).
\end{equation}

\section{Simulation of Grover's algorithm}
\label{simulation}

Grover's algorithm \cite{grover1} allows us to search an element
in an unstructured database with $N$ elements (suppose that
$N=2^n$) using $\co(\sqrt{N})$ steps. The best classical algorithm
uses $\co(N)$ steps. This quadratic speed-up is one of the
greatest successes of quantum computation so far.

Grover's algorithm has two registers, the first one with $n$
qubits and the second one with one qubit. It starts by preparing a
superposition of all computational basis states with same
amplitude. An oracle is used to probe the database (a quantum
memory is assumed to exist in some form), and it changes the sign
of the amplitude of the state which corresponds to the numerical
index of the searched element.  In order to simulate the oracle
action, we use a $n+1$ qubits generalized Toffoli gate, which
marks the state $\sket{N-1}$ (we are assuming that $N-1$ is the
searched element).

The state $\sket{N-1}$ has $n$ ones in binary notation, and that
is why the oracle just changes the sign of the amplitude of that
state. Any other state of the computational basis could be marked
as well, introducing two $X$ gates in the oracle, for each zero in
the binary representation of the state's label. The $X$ gates are
placed symmetrically with respect to the generalized Toffoli gate,
in the qubits corresponding to the zeroes in the binary
representation.

\begin{figure}
    \centering
    \psfrag{1}{1}
    \psfrag{2}{2}
    \psfrag{n}[][]{$n$}
    \psfrag{n1}[][]{$n-1$}
    \psfrag{n2}[][]{$n+1$}
    \includegraphics[]{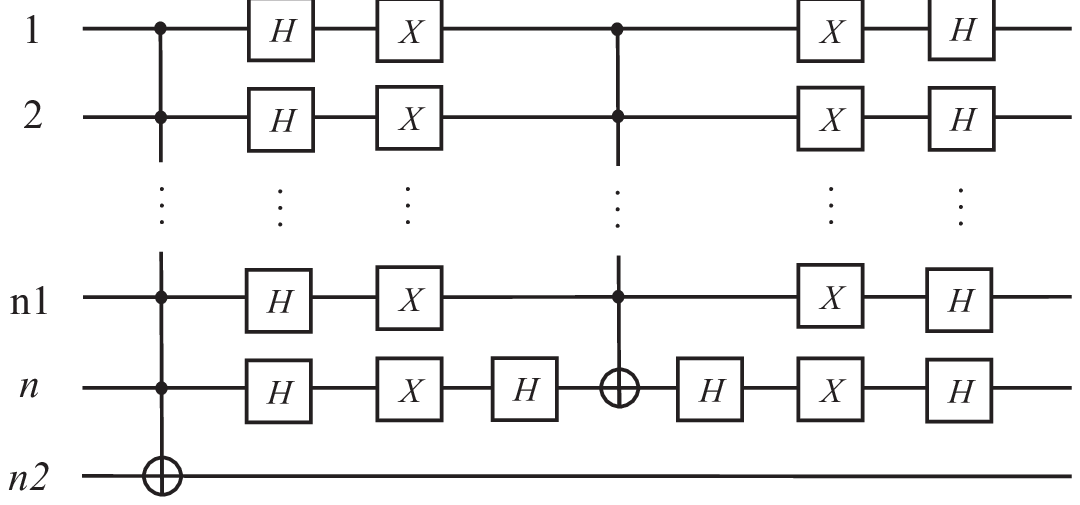}
    \caption{Grover operator.}
    \label{G}
\end{figure}

Fig. \ref{G} shows the Grover operator in terms of two generalized
Toffoli gates and one-qubit gates. This decomposition is enough to
apply the method of the previous section. Note that the method
used in \cite{cerf} would require the decomposition of the
generalized Toffoli gate into two- and one-qubit gates. This would
introduce more scalar fields ($\s$'s and $\t$'s) and lead to
cumbersome calculations. For example, for $N=4$, the method of
Cerf and Koonin would use 28 scalar fields, while ours uses 12.

A non-trivial part of our procedure is the calculation of the
decomposition (\ref{Uk}) for a general $n$-qubit gate.
Fortunately, it is easy to obtain the decomposition of the
generalized Toffoli gate:
\be
\label{decomposition}
\left(
\begin{array}{ccccccc}
1& & & & & & \\
 &1& & & & & \\
 & &\ddots & & & & \\
 & & &1& & & \\
 & & & &1& & \\
 & & & & &0&1\\
 & & & & &1&0\\
\end{array}
\right) = e^{-i \a A^{(1)} \otimes \dots \otimes A^{(n+1)} },
\end{equation}
where
\bea \a & = & \frac{\p}{2}, \non \\ A^{(1)} & = & \dots = A^{(n)}
= \frac{1-\s_z}{2} = \left(
\begin{array}{cc} 0&0\\0&1 \end{array}
\right), \\ A^{(n+1)} & = & 1 - \s_x = \left(
\begin{array}{rr} 1&-1\\-1&1
\end{array} \right) \non .
\eea

\begin{figure}
    \centering
    \psfrag{k0}[][]{$k_0$ times}
    \psfrag{1}{1}
    \psfrag{2}{2}
    \psfrag{n}[][]{$n$}
    \psfrag{n1}[][]{$n+1$}
    \includegraphics[]{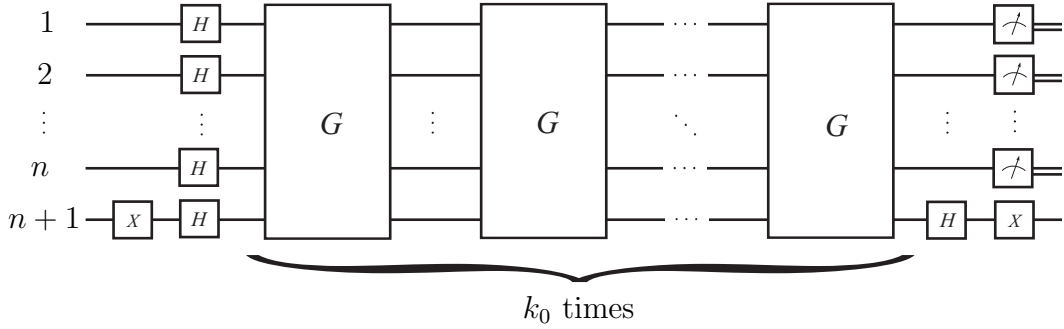}
    \caption{Full Grover algorithm.}
    \label{groveralg}
\end{figure}

The next step is the calculation of $U^{[l]}(\s, \t)$ given by
eq.(\ref{Ul}) for $l=1,\dots, n_k$. Recall that $U^{[l]}(\s, \t)$
is the ordered composition of all one-qubit gates for the qubit
$l$, including the two gates that come from the decomposition
(\ref{decomposition}), one for each generalized Toffoli gate of
Fig. \ref{groveralg}.

Since the Grover operator $G$ must be applied $k_0$ times,
we have
\be
\label{Ulk0}
U^{[l]}(\s,\t) = \left\{
\begin{array}{ll}
\left( \prod_{k=k_0}^1 U_k^{[l]}(\s,\t) \right) H & \mbox{for }
l=1,2,...,n
\\ XH \left(\prod_{k=k_0}^1 U_k^{[n+1]}(\s,\t)\right) HX &
\mbox{for } l=n+1.
\end{array}
\right.
\end{equation}
Examining fig.\ref{G}, we see that there are four
distinct kinds of $U^{[l]}(\s,\t)$, which we address now.
For $l=1$, $U^{[l]}(\s,\t)$ must be treated separately,
since from eq.(\ref{Ukl}) we see that for $l=j_1$
(here $j_1=1$) $V_k^{(1)}$ depends only on the $\s$
fields while $V_k^{(s)}$ depends on $\t_k^{(s)}$. So
\be
\label{Uk1}
U_k^{[1]} = H X \left(
\begin{array}{cc} 1 & 0 \\ 0 & e^{-i \a\; \s_{2k}^{(2)}
\dots \s_{2k}^{(n)} } \end{array} \right)
XH \left( \begin{array}{cc} 1 & 0 \\ 0 & e^{-i \a\; \s_{2k-1}^{(2)}
\dots \s_{2k-1}^{(n+1)} } \end{array} \right).
\end{equation}
For $l=2,\dots, n-1$ the gate configuration is similar, then
\be
U_k^{[l]} = H X \left(
\begin{array}{cc} 1 & 0 \\ 0 & e^{-i \a\; \t_{2k}^{(l)} }
\end{array} \right)
XH \left( \begin{array}{cc} 1 & 0 \\ 0 & e^{-i \a\; \t_{2k-1}^{(l)} }
\end{array} \right).
\end{equation}
For $l=n$, Grover operator has two extra Hadamard and a $X$
gate. Then
\be
U_k^{[n]} = \ha H X H \left(
\begin{array}{cc} 1+ e^{-2 i \a\; \t_{2k}^{(n)} }
& 1- e^{-2 i \a\; \t_{2k}^{(n)} } \\
1- e^{-2 i \a\; \t_{2k}^{(n)} }
& 1+ e^{-2 i \a\; \t_{2k}^{(n)} } \end{array} \right) H X H
\left(\begin{array}{cc} 1 & 0 \\ 0 & e^{-i \a\; \t_{2k-1}^{(n)} }
\end{array} \right).
\end{equation}
Finally, for $l=n+1$,
\be
\label{Ukn+1}
U_k^{[n+1]} = \ha \left(
\begin{array}{cc} 1 + e^{-2 i \a\; \t_{2k-1}^{(n+1)} }
& 1 - e^{-2 i \a\; \t_{2k-1}^{(n+1)} } \\
1 - e^{-2 i \a\; \t_{2k-1}^{(n+1)} }
& 1 + e^{-2 i \a\; \t_{2k-1}^{(n+1)} } \end{array} \right).
\end{equation}

Next we can insert the latter results in eq.(\ref{Ulk0}) and
obtain $U^{[l]}$. We note that in eq.(\ref{O}), $\sket {0_l}$ is
represented as $\left[\begin{array}{c} 1 \\ 0 \end{array}\right]$.
So we get for example, for observable $\co_{00} = (
\sket{0_1}\sbra{0_1} ) \otimes (\sket{0_2} \sbra{0_2})$,
\bea \label{00} \lefteqn{\co_{00}(\s,\t,\s^\prime,\t^\prime)  = }
\non \\ & & \prod_{l=1}^{n} \frac{
{{U^\dagger}^{[l]}}_{1,1}(\s^\prime,\t^\prime)
{{U}^{[l]}}_{1,1}(\s,\t) } {
{{U^\dagger}^{[l]}}_{1,1}(\s^\prime,\t^\prime)
{U^{[l]}}_{1,1}(\s,\t) +
{{U^\dagger}^{[l]}}_{1,2}(\s^\prime,\t^\prime)
{U^{[l]}}_{2,1}(\s,\t)
 }.
\eea
Similarly, in each observable we choose to test, we need to calculate
only the relevant elements of $U^{[l]}$.

In the Appendix, we describe a Maple\footnote{Maple Waterloo
Software, Inc. See http://www.maplesoft.com} program that
calculates the physical quantities of this section starting from
this point. Given $n$, the program uses eqs.(\ref{Uk1}) to
(\ref{Ukn+1}) to calculate $U^{[l]}(\s,\t)$ and
${U^\dagger}^{[l]}(\s^\prime,\t^\prime)$, and then obtains
$\co(\s,\t,\s^\prime,\t^\prime)$ for a given observable
constructed similarly to the above. Using eqs.(\ref{S}) and
(\ref{langevin}), the program obtains the Langevin equations for
the simulation of Grover's algorithm with an arbitrary number of
qubits.

The simplest case is a search space of $N=4$ elements, for which
the auxiliary fields are $\s^{(2)}_2, \s^{(2)}_1, \s^{(3)}_1,
\t^{(2)}_2, \t^{(2)}_1, \t^{(3)}_1$, and the respective primed
versions. We give below the first Langevin equation for this case.
The other equations of the system of partial differential
equations have a format similar to this one:
\bea {\frac {d\sigma^{(2)}_1}{dt\;}} & = & i{\frac{\pi}{4}}
\,\tau^{(2)}_1 + \eta \left( t \right) - {i \frac{\pi}{4}}
\,\sigma^{(3)}_1 \left( {e^{-{i\frac{\pi}{2}} \, \left(
-{\sigma^\prime}^{(2)}_2
+\sigma^{(2)}_1\sigma^{(3)}_1+\sigma^{(2)}_2 \right) }} \right.
\non \\ & &  + {e^{-{i\frac{\pi}{2}} \, \left(
-{\sigma^\prime}^{(2)}_1{\sigma^\prime}^{(3)}_1-{\sigma^\prime}^{(2)}_2
+\sigma^{(2)}_1\sigma^{(3)}_1+\sigma^{(2)}_2 \right) }}+
{e^{-{i\frac{\pi}{2}} \, \left(
-{\sigma^\prime}^{(2)}_1{\sigma^\prime}^{(3)}_1 +\sigma^{(2)}_1
\sigma^{(3)}_1 \right) }} \non \\ & & \left. -
{e^{-{i\frac{\pi}{2}} \,\sigma^{(2)}_1\sigma^{(3)}_1 }} \right)
\Bigg/ \left( {e^{-{i\frac{\pi}{2}} \, \left(
-{\sigma^\prime}^{(2)}_2+ \sigma^{(2)}_2 \right)} +
{e^{-{i\frac{\pi}{2}} \, \left( -{\sigma^\prime}^{(2)}_2
+\sigma^{(2)}_1\sigma^{(3)}_1 + \sigma^{(2)}_2 \right) }}} \right.
\non \\ & & + {e^{{i\frac{\pi}{2}} \, \left(
{\sigma^\prime}^{(2)}_1{\sigma^\prime}^{(3)}_1+{\sigma^\prime}^{(2)}_2
-\sigma^{(2)}_2 \right) }}+{e^{-{i\frac{\pi}{2}} \, \left(
-{\sigma^\prime}
^{(2)}_1{\sigma^\prime}^{(3)}_1-{\sigma^\prime}^{(2)}_2+\sigma^{(2)}_1\sigma
^{(3)}_1+\sigma^{(2)}_2 \right) }} \non \\ & &\left. +
{e^{-{i\frac{\pi}{2}} \, \left( -
{\sigma^\prime}^{(2)}_1{\sigma^\prime}^{(3)}_1
+\sigma^{(2)}_1\sigma^{(3)}_1 \right) }} +1 -
{e^{-{i\frac{\pi}{2}} \,\sigma^{(2)}_1\sigma^{(3)}_1}}-
{e^{{i\frac{\pi}{2}}
\,{\sigma^\prime}^{(2)}_1{\sigma^\prime}^{(3)}_1}} \right).
\label{langevin2} \eea
This system of equations is discretized and solved numerically.
After trying different values for the discretization parameter and
initial conditions, we plot each scalar field $\s_k^{(s)}$ as a
function of $t$ in order to determine the convergence value, to be
substituted in the previously obtained formula (\ref{00}), for the
observable $\sVEV{\co_{00}}$ and in the equivalent formulas for
$\sVEV{\co_{01}}$, $\sVEV{\co_{10}}$ and $\sVEV{\co_{11}}$,
corresponding to the observables
\be
\co_{ij} = \sket{ij}\sbra{ij} \ee
for $i,j=0,1$. Since we have chosen the oracle that changes the
sign of the state $\sket{11}\sket - $, we expect to obtain
$\sVEV{\co_{11}}$ close to 1, and the remaining ones close to 0.

Unfortunately, we are faced with technical difficulties. First,
there are some fields that do not converge to a definite value, at
least to the extent of our simulation. In our case, this issue was
circumvented by imposing the constraint that total probability
equals one, but in the general case the use of such an artifact
would not be desired. The second problem are fields that seem to
have a logarithmic behaviour. We have tried to solve this by
fitting the logarithmic behaviour by a log function, then
inserting it in the expressions and taking the limit as $t
\rightarrow \infty$. Even then, contrary to our expectation, we
get the following results
\bea \sVEV{\co_{00}} & = & .28 \non \\ \sVEV{\co_{01}} & = & .24\\
\sVEV{\co_{10}} & = & .24 \non \\ \sVEV{\co_{11}} & = & .21 \non
\eea
It is not clear to us whether we are making some mistake or this
result shows that the method does not work with Grover's
algorithm. Any comment on this issue is very welcome.

\section*{Acknowledgments}

We thank Drs. L. Davidovich and N. Zaguri and their group at UFRJ
for stimulating discussions on the subject. We are also thankful
to the Group of Quantum Computation at LNCC and to E.L. Araujo for
helping us with C implementations.

\section*{Appendix}

This appendix describes a Maple program for release 6 or higher
that calculates the Langevin equations for the simulation of
Grover's algorithm, as described in Section \ref{simulation}. The
number of qubits is fixed in the beginning of the session. The
lines beginning with the prompt \texttt{>} are Maple commands. We
make some comments about each group of commands.

\

\noindent
1. Starting a new session, defining matrices $H$, $X$ and $Z$,
and setting the number of qubits.

\begin{verbatim}
> restart;
> with(LinearAlgebra):
> H := 1/sqrt(2)*Matrix([[1,1],[1,-1]]):
> X := Matrix([[0,1],[1,0]]):
> Z := Matrix([[1,0],[0,-1]]):
> n := 2:
\end{verbatim}

\noindent
2. Function ${\texttt{k0}}$ finds the number of times the Grover operator
is applied.

\begin{verbatim}
> k0 := proc()
> local theta;
>    theta := 2*evalf(arccos(sqrt(1-1/(2^n))));
>    eval(round(Pi/(2*theta) - 1/2))
> end proc:
\end{verbatim}

\noindent 3. The next two procedures calculate
$U^{[l]}(\sigma,\tau)$ given by eq.(\ref{Ulk0}) and
$U^{[l]}_k(\sigma_k,\tau_k)$ given by eqs.(\ref{Uk1}-\ref{Ukn+1}).
They are not used directly, as we see ahead.

\begin{verbatim}
> Ul := proc(l)
>    if l<n+1 then
>       `.`(seq(Ul_k(l,k0()-s),s=0..k0()-1)).H
>    elif l=n+1 then
>       X.H.`.`(seq(Ul_k(n+1,k0()-s),s=0..k0()-1)).H.X
>    else error(`expecting l<=n+1, got %1`,l)
>    end if
> end proc:
>
> Ul_k := proc(l,k)
>    if l=1 then
>       H.X.Matrix1(mul(sigma[s,2*k][t],s=2..n)).X.H.
>             Matrix1(mul(sigma[s,2*k-1][t],s=2..n+1))
>    elif 1<l and l<n then
>       H.X.Matrix1(tau[l,2*k][t]).X.H.Matrix1(tau[l,2*k-1][t])
>    elif l=n then
>       Z.Matrix2(tau[n,2*k][t]).Z.Matrix1(tau[n,2*k-1][t])
>    elif l=n+1 then
>       Matrix2(tau[n+1,2*k-1][t])
>    end if
> end proc:
\end{verbatim}

\noindent 4. The next two procedures are auxiliary functions for
calculating the matrices of eqs.(\ref{Uk1}-\ref{Ukn+1}).

\begin{verbatim}
> Matrix1 := proc(x)
>    Matrix([[1,0],[0,exp(-I*(Pi/2)*x)]])
> end proc:
>
> Matrix2 := proc(x)
>    Matrix([[1/2+1/2*exp(-I*Pi*x), -1/2*exp(-I*Pi*x)+1/2],
>            [-1/2*exp(-I*Pi*x)+1/2, 1/2+1/2*exp(-I*Pi*x)]])
> end proc:
\end{verbatim}

\noindent 5. The next two functions calculate
$U^{[l]}(\sigma,\tau)$ and its hermitian conjugate, automatically
applying simplifying functions. An user should use these functions
since they return the simplified result.

\begin{verbatim}
> U := l -> Map(simplify,Ul(l)):
> Udagger := l -> Map(simplify, subs(I=-I, sigma=sigma1,
>                    tau=tau1, Transpose(Ul(l)))):
\end{verbatim}

\noindent
6. The next command calculates the action $S$ given by
eq.(\ref{S}), taking $\sket{0}$ as the input state for all qubits.

\begin{verbatim}
> S := - (Pi/2)*add(
>         add(sigma[s,k][t]*tau[s,k][t] -
>               sigma1[s,k][t]*tau1[s,k][t], s=2..n+2*frac(k/2)),
>                    k=1..k0()) +
>         I* add( ln( simplify( Udagger(l)[1,1]*U(l)[1,1]+
>               Udagger(l)[1,2]*U(l)[2,1] ) ),l=1..n+1):
\end{verbatim}

\noindent
7. The next commands calculate the Langevin equation given in eq.(\ref{langevin}).

\begin{verbatim}
> dS := map(factor,diff(S,sigma[2,1][t])):
> Diff(sigma[2,1][t],t) =  map(x->x*(- I/2),dS) + eta(t);
\end{verbatim}

\end{document}